\documentclass[runningheads]{llncs}
\usepackage[T1]{fontenc}
\def\doi#1{\href{https://doi.org/\detokenize{#1}}{\url{https://doi.org/\detokenize{#1}}}}
\usepackage{graphicx}
\usepackage{listings}
\usepackage{amsmath,graphicx}
\usepackage{fancyhdr}
\usepackage{bbding}
\usepackage{array}
\usepackage{xcolor}
\newcolumntype{P}[1]{>{\centering\arraybackslash}p{#1}}

\usepackage{framed,multirow}
\begin{document}

\title{Elongated Physiological Structure Segmentation via Spatial and Scale Uncertainty-aware Network}
% Spatial and Scale
% DUGl Uncertainty-Aware Network with Gated Soft Attention for Medical Image Segmentation



\author{Yinglin Zhang\inst{1,2}\and %index {First Name, Last Name}
Ruiling Xi\inst{2}\and  %index {First Name, Last Name}
Huazhu Fu\inst{4}\and  %index {First Name, Last Name}
Dave Towey\inst{1}\and
RuiBin Bai\inst{1}\and  %index {First Name, Last Name}
Risa Higashita\inst{2,3}$^\star$\and    %index {First Name, Last Name}
Jiang Liu\inst{1,2,3}$^\star$}    %index {First Name, Last Name}
% \author{}

% \footnotetext[$^\star$]{denotes corresponding author}
% \cortext[cor1]{Corresponding author}

\institute{School of Computer Science, University of Nottingham Ningbo China, Ningbo 315100, China\and
Research Institute of Trustworthy Autonomous Systems and Department of Computer Science and Engineering, Southern University of Science and Technology, Shenzhen, 518055, China \and
Tomey Corporation, Nagoya 451-0051, Japan \and
Institute of High Performance Computing (IHPC), Agency for Science, Technology and Research (A*STAR), Singapore
}
% \institute{}

\maketitle              % typeset the header of the contribution
\begin{abstract}
Robust and accurate segmentation for elongated physiological structures is challenging, especially in the ambiguous region, such as  the corneal endothelium microscope image with uneven illumination or the fundus image with disease interference. In this paper, we present a  spatial and scale uncertainty-aware network (SSU-Net) that fully uses both spatial and scale uncertainty to highlight ambiguous regions and integrate hierarchical structure contexts. First, we estimate epistemic and aleatoric spatial uncertainty maps using Monte Carlo dropout to approximate Bayesian networks. Based on these spatial uncertainty maps, we propose the gated soft uncertainty-aware (GSUA) module to guide the model to focus on ambiguous regions. Second, we extract the uncertainty under different scales and propose the multi-scale uncertainty-aware (MSUA) fusion module to integrate  structure contexts from hierarchical predictions, strengthening the final prediction. Finally, we visualize the uncertainty map of final prediction, providing interpretability for segmentation results.
Experiment results show that the SSU-Net performs best on cornea endothelial cell and retinal vessel segmentation tasks. Moreover, compared with counterpart uncertainty-based methods, SSU-Net is more accurate and robust.

\end{abstract}
\section{Introduction}
% Cornea disease is one of the leading causes of blindness \cite{2001Corneal}, and 1.8 million corneal transplants are performed worldwide \cite{2016Global}. The clinical parameters of cornea endothelium, such as cell density (CD), hexagonality (HEX), and coefficient of variation (CV), are critical evidence for cornea disease diagnosis and surgical planning. The most crucial step of clinical parameters quantification is cornea endothelium segmentation. 

Robust and accurate elongated physiological structure segmentation is crucial for computer-aided diagnosis and quantification of clinical parameters \cite{zhao2020automated} \cite{zhang2021multi}.
Manual delineation is tedious and laborious.
% and there is a subjective bias among independent clinical experts. \textcolor{purple}{Introducing clinical expert bias here maybe misleading.} 
Recently, deep learning-based methods \cite{ronneberger2015u-net} \cite{oktay2018attention} \cite{mou2021cs2} have been proposed to delineate targets automatically. However, they are not able to outline correctly in ambiguous regions where exist uneven illumination, artifacts, or interference from the disease. 
% Besides, the clinical expert needs help in understanding the reliability of the segmentation from the network. 

Many researchers have tried to use uncertainty information to concentrate on the ambiguous region, and to evaluate the reliability of model's prediction. According to the source of prediction errors \cite{gawlikowski2021survey}, uncertainty is categorized into two types: epistemic and aleatoric. The main methods for uncertainty estimation are as follows. Bayesian neural networks \cite{1994Bayesian} place a probability distribution over  model weights, but are hard to optimize. Monte Carlo dropout \cite{leibig2017leveraging} approximates the Gaussian process by embedding the dropout operation into the neural network layers and calculating the variance of several times inference. Deep Ensembles \cite{lakshminarayanan2017simple} combine the outputs from a group of independent models to estimate uncertainty.
Softmax uncertainty \cite{pidaparthy2021automatic} \cite{pearce2021understanding} \cite{mehrtash2020confidence} performs well in distinguishing examples that are easy or fallible to classify. 
% Introducing perturbation to input data produced a number of reasonable hypotheses that also capture the uncertainty of segmentation \cite{kohl2018probabilistic}.
Once the uncertainty information has been estimated, we are able to pay more attention to the ambiguous region. 
% and obtain the confidence of segmentation. 
Xie et al. \cite{xie2022uncertainty} used the cross-attention module to extract influential features for ambiguous regions based on pixel-level uncertainty. 
Yang et al. \cite{yang2022uncertainty} achieved uncertainty awareness by training with a multi-confidence mask, and further used self-attention block with feature aware filter together to highlight uncertain areas.
Wang et al. \cite{wang2021medical} annotated alpha matte for medical images and used it as a soft label to intuitively promote the network to focus on uncertain areas.
Kohl et al. \cite{kohl2018probabilistic} proposed a generative model to produce multiple reasonable hypotheses for clinical experts to select from, which improved the diagnosis reliability. 
% Mehrtash et al. \cite{mehrtash2020confidence} proposed decision confidence by averaging the pixel-level entropy value across different classes in the prediction.
However, existing works applied the 'hard' attention to utilize uncertainty, which lacks the ability of adaptive adjustment and ignores neighboring uncertain regions.
% and only considered the pixel-level uncertainty at a single scale. 
In addition, features at different scales contain rich structural and semantic contexts, which are essential for elongated physiological structure segmentation, such as cobweb corneal endothelial cells and retinal vessels.
% medical image segmentation

This paper proposes a spatial and scale uncertainty-aware network (SSU-Net) for elongated physiological structure segmentation, which fully uses both spatial and scale uncertainty to highlight ambiguous regions and integrate hierarchical structure contexts. 
% First, we obtain epistemic and aleatoric spatial uncertainty maps using Monte Carlo dropout to approximate Bayesian networks. 
First, we use a gated soft uncertainty-aware (GSUA) module to adaptively highlight ambiguous areas based on spatial uncertainty maps. Second, we extract the uncertainty under different scales and propose the multi-scale uncertainty-aware (MSUA) fusion module to integrate hierarchical predictions for enhancing the final segmentation. Experiment results on segmentation tasks of the cornea endothelium and retinal vessel show the effectiveness of SSU-Net.  

% Finally, we visualized the output of the uncertainty map at different scales in the form of a heatmap, providing interpretability for segmentation results.

% uses a gated soft attention module to highlight ambiguous regions based on spatial uncertainty and the SUF module to strengthen the final segmentation by fusing predictions under multiple scales.  obtained by Monte Carlo approximation Bayesian sampling, we use the gated soft attention (GSA) module to guide the model to focus on the ambiguous spatial region. Then, we propose a multi-scale fusion module based on uncertainty information at different scales for segmentation.

% Kohl et al [xx] proposed a probabilistic generative model to capture the uncertainty by introducing perturbation of inputs. 

\section{Method}
\begin{figure*}[t]
\includegraphics[width=\textwidth]{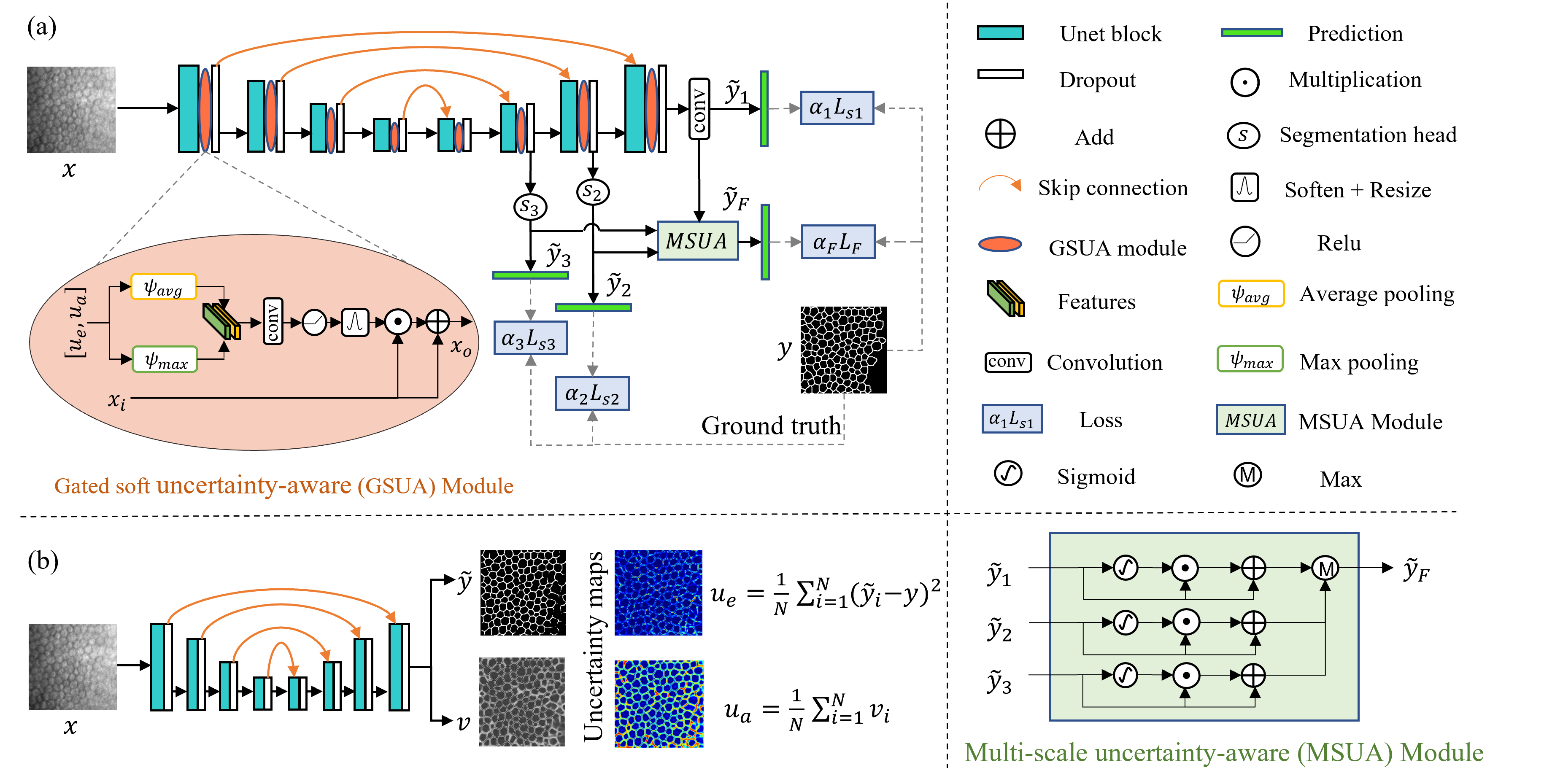}
\caption{The pipeline of the proposed algorithm. (a) The framework of spatial and scale uncertainty-aware network, SSU-Net. (b) We estimate the spatial uncertainty maps with Bayesian approximate network. 
% In stage one, we calculate the epistemic and aleatoric spatial uncertainty maps with the Monte Carlo dropout method. In stage two, we use the GSA module to highlight the ambiguous region and integrate multi-scale predictions with the MPF module.
} \label{fig1}
\end{figure*}

Fig.\ref{fig1} (a) illustrates the framework of the proposed spatial and scale uncertainty-aware network, SSU-Net. The gated soft uncertainty-aware (GSUA) module enables the network to focus on the ambiguous region indicated by the spatial uncertainty maps $[u_{e},u_{a}]$. Specifically, we construct a Bayesian approximate network to generate spatial uncertainty maps by introducing Monte Carlo dropout \cite{dropout2016} into U-Net, as shown in Fig.\ref{fig1} (b). The Bayesian approximate network has two outputs: segmentation prediction $\hat{y}$ and the estimation of aleatoric uncertainty $v$. We can calculate the epistemic and aleatoric uncertainty maps, $u_{e}$ and $u_{a}$, after multiple inferences. Furthermore, we consider the sigmoid probabilities of predictions under different scales as the second uncertainty source, and fuse the predictions $\{\hat{y}_{1}, \hat{y}_{2}, \hat{y}_{3}\}$ from multiple scales using the multi-scale uncertainty-aware (MSUA) module. $\hat{y}_{F}$ is the final target output.

% \subsection{Uncertainty Estimation}
% In this work, we extract spatial and scale uncertainty to highlight ambiguous regions and integrate multi-scale predictions.

\subsection{Spatial Uncertainty and Gated Soft Uncertainty-aware Module}

\textbf{Spatial Uncertainty}. Since the epistemic and aleatoric uncertainty maps are used to find the hard-to-classify spatial areas in this work, we regard them as spatial uncertainty.
Referring to \cite{kendall2017uncertainties}, we add dropout after each UNet block to approximate the Bayesian network, which learns the segmentation $\tilde{y}$ and aleatoric uncertainty $v$ simultaneously. During inference, we sample a group of predictions $\{\tilde{y}_{i}\}^{N}_{i=1}$ and $\{v_{i}\}^{N}_{i=1}$ by $N$ stochastic forward pass. In this work, we set $N=16$. The epistemic $u_{e}$ and aleatoric $u_{a}$ uncertainty are formulated by Equation (\ref{eq1}), where $y$ is the ground truth.
\begin{equation}\label{eq1}
    \begin{split}
    &u_{e} = \frac{1}{N} {\textstyle \sum_{i=1}^{N}(\tilde{y}_{i}-y)^{2}}, u_{a} = \frac{1}{N} {\textstyle \sum_{i=1}^{N}v_{i}}   
    \end{split}
\end{equation}

\noindent\textbf{Gated Soft Uncertainty-aware Module}.
To endow the uncertainty-aware module with adaptive adjustment ability, we propose the gated soft uncertainty-aware (GSUA) module, as illustrated in Fig.\ref{fig1}. We extract salient descriptions from uncertainty maps by two parallel pooling $[\psi_{avg}, \psi_{max}]$ and a $1\times1$ convolution $f(\cdot)$ operation. The $relu$ operation is set as a switch to filter out areas with small uncertainty values, further strengthening our attention on areas with high uncertainty. Since it is also usually difficult to classify the area adjacent to the high-uncertainty regions, we use the Gaussian kernel to soften the boundary in such regions. The GSUA module is formulated by:  

\begin{equation}\label{eq2}
x_{o}=x_{i} \odot  g_{s}(\sigma (relu(f([\psi_{avg}(\mathbf{u}),  \psi_{max}(\mathbf{u})]))))+x_{i}    
\end{equation}
where $x_{i}, x_{o}\in R^{N\times c\times h\times w}$ are the input and output features respectively; $\mathbf{u}=[u_{a},u_{e}] \in R^{N\times 2\times H\times W}$ is a tensor of uncertainty maps; $\psi_{avg}$ and $\psi_{max}$ represent average and max pooling; $\sigma$ is the sigmoid function; $g_{s}$ denotes a convolution operation with Gaussian kernel and resizes the attention maps to the size of input features; $\odot$ is element-wise multiplication.
% Finally, we resized the attention map and multiplied it with the input feature $x_{in}$, and add it to $x_{in}$. 

\subsection{Scale Uncertainty and Multi-scale Uncertainty-aware Module}
\textbf{Scale Uncertainty}. To integrate the predictions from hierarchical layers during model training, we capture the uncertainty under multiple scales. The sigmoid function is a simple and effective way to estimate uncertainty for the binary classification task. We extract the multi-scale uncertainty by Equation (\ref{eq3}), where $u_{s}$ is the uncertainty map of prediction $\Tilde{y}_{s}$ under scale $s \in \{1,2,3\}$.
\begin{equation}\label{eq3}
u_{s} = \frac{1}{1+e^{-\tilde{y}_{s}}}
\end{equation}

\noindent\textbf{Multi-scale Uncertainty-aware Module}. With the uncertainty maps from different scales, all the hierarchical predictions $\{\Tilde{y}_{1}, \Tilde{y}_{2}, \Tilde{y}_{3}\}$ are fused by the MSUA module to generate the enhanced prediction $\Tilde{y}_{F}$, as illustrate in Fig.\ref{fig1}. 
The uncertainty map $u_{s}$ provides the classification confidence for each pixel. Therefore, we use $u_{s}$ to highlight the confident region of $\Tilde{y}_{s}$ and further extract the max value across the different scales. The process is formulated by:
\begin{equation}\label{eq4}
\tilde{y}_{F}(i,j)=\underset{s \in \{1,2,3\}}{max}(y_{s}(i,j) \odot \sigma (y_{s}(i,j))+y_{s}(i,j))
\end{equation} 
where $\tilde{y}_{F}(i,j)$ denotes the pixel value at location $(i,j)$ of enhanced prediction; $y_{s}(i,j)$ is the value of prediction under scale $s \in \{1,2,3\}$; and $\sigma$ denotes the sigmoid operation of Equation (\ref{eq3}).

\subsection{Objective Function}
As shown in Fig.\ref{fig1}, we optimize the model with supervision on four segmentation branches simultaneously, including supervision for predicting three scales and the final enhanced output. The loss function is summarized as follows: 

\begin{equation}\label{eq5}
\mathcal{L}_{total}=\alpha_{1}\mathcal{L}_{s1}+\alpha_{2}\mathcal{L}_{s2}+\alpha_{3}\mathcal{L}_{s3}+\alpha_{F}\mathcal{L}_{F}
\end{equation}
where $\alpha_{1}, \alpha_{2}, \alpha_{3}$, and $\alpha_{F}$ are the weight parameters for sub loss $\mathcal{L}_{s1}, \mathcal{L}_{s2}, \mathcal{L}_{s3}$, and  $\mathcal{L}_{F}$. In this experiment, we set all the weight parameters as 1. For these sub-losses, we adopt binary cross-entropy loss, as shown in Equation (\ref{eq6}).

\begin{equation}\label{eq6}
\mathcal{L}=-[ylog\Tilde{y}+(1-y)log(1-\Tilde{y})] 
\end{equation}
where $y$ is the ground truth; the positive class value of each pixel is 1; the negative class value is 0; and $\Tilde{y} \in (0,1)$ is the predicted probability value.

\section{Experiment}
% In this section, we investigated the effectiveness of core components: gated soft attention (GSA), and multi-scale predictions fusion (MPF) modules. We also compared the proposed DUGl uncertainty aware network (SSU-Net) with the state-of-the-art (SOTA) methods on the cornea endothelial cell and retinal vessel segmentation tasks.

\subsection{Dataset}
% Three datasets are used in this work.
Two cornea endothelium microscope image  datasets, TM-EM3000 and Rodrep, and one retinal fundus image dataset, FIVES, are used in this work. 
The private dataset \textbf{TM-EM3000} contains 183 images measured by EM3000 specular microscope (Tomey Corporation, Japan). Following Ruggeri et al. \cite{ruggeri2010system}, we cropped a $192 \times 192$ pixels sub-region from its $260 \times 480$ pixels whole image. We used 155 images for model training, ten images for validation, and 18 images for testing. \textbf{Rodrep} \cite{selig2015fully} contains 52 in-vivo confocal corneal microscope images, 
% sampled by a white light slit scanning confocal microscope. These images 
from 23 Fuchs patients with endothelial corneal dystrophy. We used 40 for training, five images for validation, and seven for testing. \textbf{FIVES} \cite{jin2022fives} is the largest known high-resolution fundus image dataset: It covers normal eyes and three different eye diseases with a balanced distribution. 
There are 800 high-resolution images and the corresponding manual annotations, with 550 for training, 50 for validation, and 200 for testing.

\subsection{Evaluation Metrics and Implementation Details}
% \noindent\textbf{Evaluation Metrics}.
% For the binary segmentation task of cornea endothelial cells and retinal vessels, 
There is a class imbalance between foreground and background pixels. To better evaluate the segmentation performance, we choose $Dice$ score \cite{taha2015metrics}, $mIoU$, and $mAcc$ as evaluation metrics.
% , as follows.
% \begin{equation}\label{eq7}
% %\begin{split}
%     mIoU=\frac{1}{T}{\textstyle \sum_{i=1}^{T}}\frac{tp_{i}}{fp_{i}+tp_{i}+fn_{i}},  mAcc=\frac{1}{T}{\textstyle \sum_{i=1}^{T}} \frac{tp_{i}}{tp_{i}+fn_{i}}
% %\end{split}
% \end{equation}
% where $tp_{i}$ and $fp_{i}$ are the number of positive pixels classified correctly and misclassified of $i_{th}$ class; $tn_{i}$ and $fn_{i}$ correspond to negative pixels, and $T$ is the number of classes.
% \noindent\textbf{Implementation Details}. 
We optimized the models using the RMSprop strategy with momentum = 0.9 and weight decay = 1e-8 for 100 epochs. The initial learning rate was 2e-4, and the input size of all networks was uniformly set to $256 \times 256$. Random shift and rescaling within a range of $[-0.3, +0.3]$ were used for data augmentation. We set the batch size to 1 based on our empirical observations. For uncertainty-based models, we set the dropout rate as $0.5$ and no data augmentation. During testing, we inferred $N=16$ times and obtained the final prediction $\bar{y} =  \frac{1}{N}{\textstyle \sum_{i=1}^{N}} \Tilde{y}_{F}^{i}$ and the epistemic uncertainty $u'_{e}= \frac{1}{N}{\textstyle \sum_{i=1}^{N}}(\Tilde{y}_{F}^{i} - \bar{y})^{2}$.

\subsection{Ablation Study}
We investigated the influence of GSUA and MSUA modules on TM-EM3000, as shown in Table \ref{tab1}. 
% Variant1 is the baseline model, constructed with only the Unet block and dropout layers. For Variant2, we added the MPF module and corresponding multi-scale supervision branch to the baseline. For Varinat3, we added the GSA module between the Unet and the dropout layer to the baseline.
% According to the ablation study results,
The MSUA increased performance by 0.69\% on the Dice score, and GSUA increased by 0.19\%. The MSUA module brought more improvement than GSUA, which indicated that multi-scale context is crucial for cornea endothelium cell segmentation. 
When using both GSUA and MSUA modules simultaneously, we achieved the best performance. 
% with a Dice score of 77.16\%, mIoU of 77.02\%, and mAcc of 87.34\%. 

\begin{table}[htb]
% \vspace{-2em}
%\setlength{\abovecaptionskip}{-0.5cm}
\caption{Ablation study on TM-EM3000. GSUA denotes the gated soft uncertainty-aware module. MSUA means the multi-scale uncertainty-aware fusion module.}\label{tab1}
\begin{center}
\begin{tabular*}{\linewidth}{P{2.0cm}| P{1.75cm} P{1.75cm} P{2.0cm} P{2.0cm} P{2.0cm}} %{lll} 
\hline
% \multirow{2}{*}{\textbf{TR}}   &\multirow{2}{*}{\textbf{B-E}}   
% &\multicolumn{2}{c}{\textbf{DICE}} \\
\textbf{Models}    &\textbf{GSUA}    &\textbf{MSUA}       
&\textbf{Dice(\%)} $\uparrow$     &\textbf{mIoU(\%)} $\uparrow$     &\textbf{mAcc(\%)} $\uparrow$ \\
\hline
Variant1    &\XSolidBrush    &\XSolidBrush       &76.04  &76.32  &85.52  \\
Variant2    &\XSolidBrush    &\Checkmark         &76.73  &76.68  & \textbf{87.38} \\
Variant3    &\Checkmark      &\XSolidBrush       &76.23  &76.42  &86.08  \\
Variant4    &\Checkmark      &\Checkmark         &\textbf{77.16}  &\textbf{77.02}  &87.34  \\
\hline
\end{tabular*}
\end{center}
\end{table}
\vspace{-1.0cm}

\begin{figure}
\includegraphics[width=\textwidth]{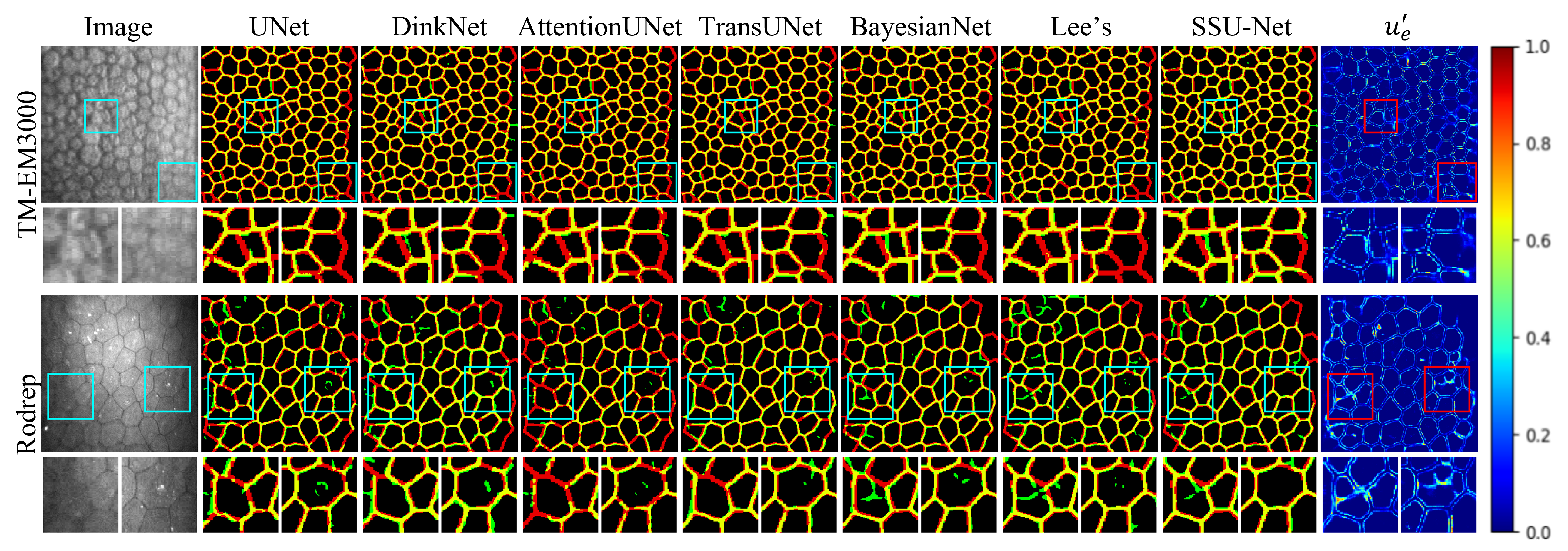}
\caption{Visualization for cornea endothelial cell segmentation. Red, green, and yellow line presents manual label, predicted segmentation result, and their overlap region. Indicated by the uncertainty map $u'_{e}$ of the SSU-Net , we zoomed in two ambiguous local regions for clear observation. } \label{fig3}
\end{figure}

\subsection{Comparison with State-of-the-art Methods}
To study the effectiveness of the proposed SSU-Net, we compared it with a series of state-of-the-art methods.
On TM-EM3000 and Rodrep, we implemented several popular networks for comparison: UNet \cite{ronneberger2015u-net}, D-LinkNet \cite{zhou2018d}, AttentionUNet \cite{oktay2018attention}, TransUNet \cite{chen2021transunet}, and uncertainty-based counterparts,
Monte Carlo (MC) BayesianNet \cite{gal2016dropout}, Lee's method \cite{lee2022method}. On the fundus image dataset FIVES, we additionally implemented several recent retinal vessel segmentation algorithms: FR-UNet \cite{liu2022full}, SA-UNet \cite{guo2021sa}, and IterNet \cite{li2020iternet}.

\begin{table}[htb]
% \vspace{-2em}
%\setlength{\abovecaptionskip}{-0.5cm}
\caption{Comparison of SSU-Net with some SOTA methods on cornea endothelial cell and retinal vessel segmentation tasks.}\label{tab2}
\begin{center}
\begin{tabular*}{\linewidth}{P{2.0cm}| P{3.5cm}| P{2.0cm} P{2.0cm} P{2.0cm}} %{lll} 
\hline
% \multirow{2}{*}{\textbf{TR}}   &\multirow{2}{*}{\textbf{B-E}}   
% &\multicolumn{2}{c}{\textbf{DICE}} \\
\textbf{Dataset}    &\textbf{Models}      
&\textbf{Dice (\%)} $\uparrow$     &\textbf{mIoU (\%)} $\uparrow$     &\textbf{mAcc (\%)} $\uparrow$ \\
\hline
\multirow{7}{*}{TM-EM3000}   &UNet       &71.28  &72.67  &81.34  \\
% &UNet++         &73.31  &74.28  &82.68 \\
&D-LinkNet       &72.71  &73.60  &82.67  \\
&AttentionUNet         &72.29  &73.46  &82.06  \\
&TransUNet         &71.65  &72.78  &82.20  \\
&BayesianNet    &75.42    &75.71  &85.66  \\
&Lee's         &76.40  &76.53  &85.80  \\
&\textbf{SSU-Net}         &\textbf{77.16}  &\textbf{77.02}  &\textbf{87.34}  \\
\hline
\multirow{7}{*}{Rodrep}   &UNet       &64.89  &67.89   &78.50   \\
% &UNet++         &  &  &  \\
&D-LinkNet       &65.40  &68.22   &79.21    \\
&AttentionUNet         &60.55  &65.68   &74.45    \\
&TransUNet         &66.56  &68.87   &80.43    \\
&BayesianNet    &65.62   &68.15   &80.24  \\
&Lee's         &63.51  &66.62  &79.44  \\
&\textbf{SSU-Net}       &\textbf{68.49}   &\textbf{70.10} &\textbf{82.58}   \\
\hline
\multirow{10}{*}{FIVES}   &UNet       &78.99  &82.59  &86.05  \\
% &UNet++         &  &  &  \\
&D-LinkNet       &73.45  &78.03  &82.89  \\
&AttentionUNet         &78.09  &81.82  &85.26  \\
&TransUNet         &80.81  &83.19  &88.18  \\
&FR-UNet        &78.47 &82.09  &85.51  \\
&SA-UNet         &79.15  &82.05  &88.27  \\
&IterNet        &79.32  &82.54   &85.56  \\
&BayesianNet  &88.70   &89.65   &93.01   \\
&Lee's         &88.79  &89.70  &93.37  \\
&\textbf{SSU-Net}         &\textbf{89.07}  &\textbf{89.88}  &\textbf{93.56}  \\
\hline
\end{tabular*}
\end{center}
\end{table}
% \vspace{-0.5cm}

\begin{figure*}[t]
\includegraphics[width=\textwidth]{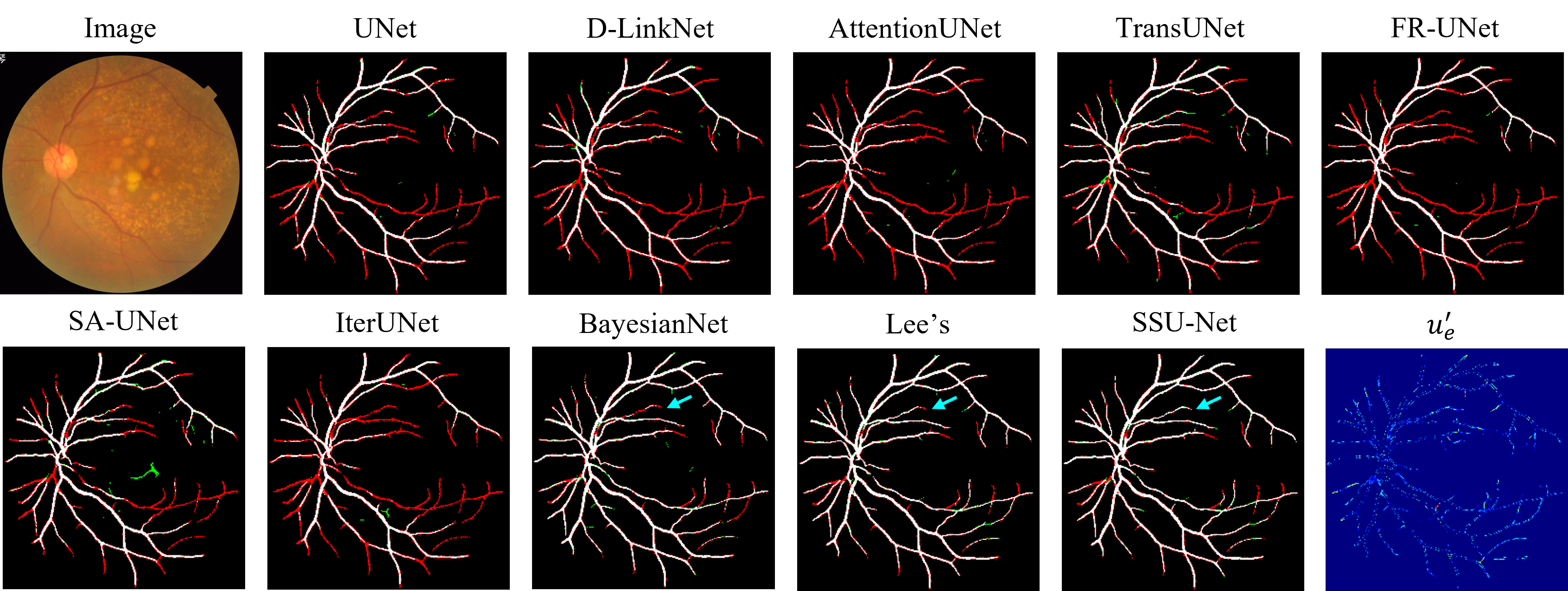}
\caption{Visualization results for retinal vessel segmentation on FIVES. Red, green, and yellow line presents manual label, predicted segmentation result, and their overlap region. The cyan arrow points to a sub-region with differences among uncertainty-based methods. We also visualize the epistemic uncertainty map $u'_{e}$ of our SSU-Net.} \label{fig4}
\end{figure*}

As shown in Table \ref{tab2}, the proposed SSU-Net achieved the best performance.
On the TM-EM3000 dataset, the uncertainty-based methods outperformed the typical convolution and attention methods, which proves that introducing uncertainty is beneficial.
% information enhances the robustness and accuracy of the model. 
% The SSU-Net increased the Dice score by $0.76\%$, mIoU by $0.49\%$, and mAcc by $1.54\%$ than Lee's method. 
% As observed in the first row of Fig.\ref{fig3}, the qualitative result also suggests that uncertainty-based methods effectively improved the segmentation performance in ambiguous regions. 
On the Rodrep dataset, SSU-Net performed considerably better than Lee's uncertainty method, improving the Dice score by $4.98\%$, mIoU by $3.48\%$, and mAcc by $3.14\%$. The results further suggest that the multi-scale predictions fusion module is crucial to elevate the robustness. 
According to the indication of uncertainty map $u'_{e}$, we cropped and zoomed in two ambiguous regions of each image, as shown in Fig.\ref{fig3}.
The visualization results suggested that the proposed SSU-Net effectively improved the segmentation performance in ambiguous regions. 
% The qualitative analysis is shown in the second row of Fig.\ref{fig3}. 
On the FIVES dataset, the performance of the specialized network for retinal blood vessel segmentation in the fundus was similar to that of UNet, TransUNet, and AttettionUNet. The uncertainty-based methods are uniformly significantly superior to the above methods. The proposed SSU-Net achieved the best performance, increasing the Dice score by $10.08\%$, mIoU by $7.29\%$, and mAcc by $7.51\%$ compared with UNet. Qualitative analysis is shown in Fig.\ref{fig4}, further supporting the conclusions of quantitative analysis.

% \begin{table}[htb]
% % \vspace{-2em}
% %\setlength{\abovecaptionskip}{-0.5cm}
% \caption{Compare SSU-Net with other SOTA methods on retinal vessel segmentation task.}\label{tab3}
% \begin{center}
% \begin{tabular*}{\linewidth}{P{2.0cm}| P{3.5cm}| P{2.0cm} P{2.0cm} P{2.0cm}} %{lll} 
% \hline
% % \multirow{2}{*}{\textbf{TR}}   &\multirow{2}{*}{\textbf{B-E}}   
% % &\multicolumn{2}{c}{\textbf{DICE}} \\
% \textbf{Dataset}    &\textbf{Models}      
% &\textbf{Dice (\%)} $\uparrow$     &\textbf{mIoU (\%)} $\uparrow$     &\textbf{mAcc (\%)} $\uparrow$ \\
% \hline
% \multirow{8}{*}{FIVES}   &UNet       &78.99  &82.59  &86.05  \\
% % &UNet++         &  &  &  \\
% &D-LinkNet       &73.45  &78.03  &82.89  \\
% &AttentionUNet         &78.09  &81.82  &85.26  \\
% &TransUNet         &80.81  &83.19  &88.18  \\
% &FR-UNet        &78.47 &82.09  &85.51  \\
% &SA-UNet         &79.15  &82.05  &88.27  \\
% &IterNet        &79.32  &82.54   &85.56  \\
% &BayesianNet  &88.70   &89.65   &93.01   \\
% &Lee's         &88.79  &89.70  &93.37  \\
% &\textbf{SSU-Net}         &\textbf{89.07}  &\textbf{89.88}  &\textbf{93.56}  \\
% \hline
% \end{tabular*}
% \end{center}
% \end{table}

% \vspace{-1.0cm}

\section{Conclusion}
This paper proposes a spatial and scale uncertainty-aware network (SSU-Net) for elongated physiological structure segmentation.
% We focus on the spatial uncertain region with the soft gated attention (GSA) module and enhance the final prediction with the multi-scale predictions fusion (MPF) module. 
The ablation study shows the effectiveness of core components: the soft gated uncertainty-aware (GSUA) and the multi-scale uncertainty-aware (MSUA) fusion modules. Compared with some SOTA methods on cornea endothelial cell and retinal vessel image segmentation tasks, the proposed SSU-Net achieved the best segmentation performance and is more robust than other uncertainty-based methods. It is noteworthy that the SSU-Net performed considerably better than specialized retinal vessel segmentation networks.
In the future, we plan to conduct experiments on various challenging situations to further explore the characteristics of SSU-Net.

% \subsubsection{Acknowledgements} Please place your acknowledgments at
% the end of the paper, preceded by an unnumbered run-in heading (i.e. 3rd-level heading).

%\linespread{1.15} %% add by cai
\bibliographystyle{splncs04}
\bibliography{ref}

\end{document}